\documentclass[preprint,aps,prc,showpacs,nofootinbib]{revtex4}%
\usepackage{amssymb}
\usepackage{epsfig}
\usepackage{amsmath,graphicx}
\usepackage{amsmath}
\usepackage{amsfonts}
\usepackage{graphicx}%
\providecommand{\U}[1]{\protect\rule{.1in}{.1in}}
\providecommand{\U}[1]{\protect\rule{.1in}{.1in}}

\newcommand\ba{\begin{eqnarray}}
\newcommand\ea{\end{eqnarray}}

\begin{document}
\title{On mass corrections to the decay $P\rightarrow l^{+}l^{-}$ }
\author{A. E. Dorokhov, M. A. Ivanov}
\affiliation{\textit{JINR-BLTP, 141980 Dubna, Moscow region, Russian Federation}}
\date{\today }

\begin{abstract}
We use the Mellin-Barnes representation in order to improve the theoretical
estimate of mass corrections to the width of light pseudoscalar meson decay
into a lepton pair, $P\rightarrow l^{+}l^{-}$ . The full resummation of the
terms $\ln(m_{l}^{2}/\Lambda^{2})\left(  m_{l}^{2}/\Lambda^{2}\right)  ^{n}$
and $\left(  m_{l}^{2}/\Lambda^{2}\right)  ^{n}$ to the decay amplitude is
performed, where $m_{l}$ is the lepton mass and $\Lambda\approx m_{\rho}$ is
the characteristic scale of the $P\rightarrow\gamma^{\ast}\gamma^{\ast}$ form
factor. The total effect of mass corrections for the $e^{+}e^{-}$ channel is
negligible and for the $\mu^{+}\mu^{-}$ channel its order is of a few per cent.

\end{abstract}
\maketitle


\section{Introduction}

Rare decays of mesons serve as a low-energy test of the Standard Model.
Accuracy of experiments has increased significantly in recent years.
Theoretically, one of the main limitations comes from the large distance
contributions of the strong sector of the Standard Model where the
perturbative QCD theory does not work. However, in some important cases the
result can be essentially improved by relating these poorly known
contributions to other experimentally known processes. The famous example is
the calculation of the hadronic vacuum polarization contribution to the
anomalous magnetic moment of muon $\left(  g-2\right)  _{\mu}$ where the data
of the processes $e^{+}e^{-}\rightarrow hadrons$ and $\tau\rightarrow hadrons$
are essential to reduce the uncertainty (see for review
\cite{Miller:2007kk,Passera:2007fk,Dorokhov:2005ff,Jegerlehner:2007xe}). It
turns out that this is also the case for the rare decays of light pseudoscalar
mesons into a lepton pair \cite{Dorokhov:2007bd}. Interest in these processes
revived after new precise measurement of the decay $\pi_{0}\rightarrow
e^{+}e^{-}$ by the KTeV collaboration \cite{Abouzaid:2007md}. The Standard
Model prediction \cite{Dorokhov:2007bd} disagrees with the KTeV measurement by
$3.3\sigma$, with the theoretical accuracy exceeding the experimental one.
\begin{figure}[th]
\includegraphics[width=5cm]{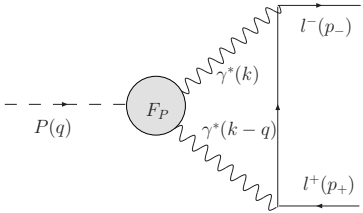}\caption{Triangle diagram for the
$P\rightarrow l^{+}l^{-}$ process with the pseudoscalar meson form factor
$P\rightarrow\gamma^{\ast}\gamma^{\ast}$ in the vertex.}%
\label{fig:triangle}%
\end{figure}

In the lowest order of QED perturbation theory, the photonless decay of the
neutral meson, $P(q)\rightarrow l^{-}(p_{-})+l^{+}(p_{+}),\quad q^{2}%
=M^{2},\quad p_{\pm}^{2}=m^{2},$ ($M$ meson mass, $m$ lepton mass) is
described by the one-loop Feynman amplitude (Fig. \ref{fig:triangle})
corresponding to the conversion of the meson through two virtual photons into
a lepton pair. The normalized branching ratio is given by
\cite{Dr59,BG60,Bergstrom:1982zq}
\begin{equation}
R_{0}(P\rightarrow l^{+}l^{-})=\frac{B_{0}\left(  P\rightarrow l^{+}%
l^{-}\right)  }{B\left(  P\rightarrow\gamma\gamma\right)  }=2\beta\left(
M^{2}\right)  \left(  \frac{\alpha m}{\pi M}\right)  ^{2}|\mathcal{A}\left(
M^{2}\right)  |^{2}, \label{Bpi}%
\end{equation}
where $\beta\left(  q^{2}\right)  =\sqrt{1-4m_{l}^{2}/q^{2}}$ and the reduced
amplitude is
\begin{equation}
\mathcal{A}\left(  q^{2}\right)  =\frac{2}{q^{2}}\int\frac{d^{4}k}{i\pi^{2}%
}\frac{(qk)^{2}-q^{2}k^{2}}{(k^{2}+i\epsilon)\left[  (q-k)^{2}+i\epsilon
\right]  \left[  (p_{-}-k)^{2}-m^{2}+i\epsilon\right]  }F_{P\gamma^{\ast
}\gamma^{\ast}}(-k^{2},-(q-k)^{2}), \label{Rq}%
\end{equation}
with the transition form factor $F_{P\gamma^{\ast}\gamma^{\ast}}(-k^{2}%
,-q^{2})$ being normalized as $F_{P\gamma^{\ast}\gamma^{\ast}}(0,0)=1$.

The imaginary part of the on-shell amplitude $\mathcal{A}\left(  q^{2}%
=M^{2}\right)  $ comes from the contribution of real photons in the
intermediate state and can be found in a model independent way \cite{BG60}%
\begin{equation}
\mathrm{Im}\mathcal{A}(M^{2})=\frac{\pi}{2\beta\left(  M^{2}\right)  }%
\ln\left(  y\left(  M^{2}\right)  \right)  ,\qquad y\left(  q^{2}\right)
=\frac{1-\beta\left(  q^{2}\right)  }{1+\beta\left(  q^{2}\right)  }.
\label{ImR}%
\end{equation}
A once-subtracted dispersion relation for the amplitude in Eq.~(\ref{Rq}) is
written as\footnote{In this derivation it is tacitly assumed that the
imaginary part of the off-shell amplitude $\mathcal{A}(q^{2})$ has the same
form as in (\ref{ImR}) with $M^{2}$ substituted by $q^{2}$.}
\cite{Bergstrom:1983ay}
\begin{equation}
\mathcal{A}\left(  q^{2}\right)  =\mathcal{A}\left(  q^{2}=0\right)
+\frac{q^{2}}{\pi}\int_{0}^{\infty}ds\frac{\operatorname{Im}\mathcal{A}\left(
s\right)  }{s\left(  s-q^{2}\right)  }. \label{DispRel}%
\end{equation}
The second term in Eq.~(\ref{DispRel}) takes into account strong $q^{2}$
dependence of the amplitude around the point $q^{2}=0$ occurring due to the
branch cut coming from the two-photon intermediate state. Integrating
Eq.~(\ref{DispRel}) for $q^{2}\geq4m_{e}^{2}$ one arrives at
\cite{D'Ambrosio:1986ze,Savage:1992ac,Ametller:1993we}%
\begin{equation}
\operatorname{Re}\mathcal{A}\left(  q^{2}\right)  =\mathcal{A}\left(
q^{2}=0\right)  +\frac{1}{\beta\left(  q^{2}\right)  }\left[  \frac{1}{4}%
\ln^{2}\left(  y\left(  q^{2}\right)  \right)  +\frac{\pi^{2}}{12}%
+\mathrm{Li}_{2}\left(  -y\left(  q^{2}\right)  \right)  \right]  ,
\label{ReR}%
\end{equation}
where $\mathrm{Li}_{2}\left(  z\right)  =-\int_{0}^{z}\left(  dt/t\right)
\ln\left(  1-t\right)  $ is the dilogarithm function.

Usually, the subtraction constant in (\ref{ReR}), containing the nontrivial
dynamics of the process, is calculated within different models describing the
form factor $F_{P\gamma^{\ast}\gamma^{\ast}}(k^{2},q^{2})$ (\textit{e.g.}
\cite{Bergstrom:1983ay,Savage:1992ac,Dorokhov:2007bd}). However, it has
recently been shown in \cite{Dorokhov:2007bd} that this constant may be
expressed in terms of the inverse moment of the transition form factor given
in symmetric kinematics of spacelike photons, $G(t)\equiv F_{P\gamma^{\ast
}\gamma^{\ast}}\left(  t,t\right)  ,$%
\begin{equation}
\mathcal{A}^{0}\left(  q^{2}=0\right)  =3\ln\left(  \frac{m}{\mu}\right)
-\frac{3}{2}\left[  \int_{0}^{\mu^{2}}dt\frac{G(t)-1}{t}+\int_{\mu^{2}%
}^{\infty}dt\frac{G\left(  t\right)  }{t}\right]  -\frac{5}{4}. \label{R0}%
\end{equation}
Here, $\mu$ is an arbitrary (factorization) scale. One should note that the
logarithmic dependence of the first term on $\mu$ is compensated by the scale
dependence of the integrals in the brackets.

The accuracy of these calculations is determined by omitted small power
corrections of the order $O(\frac{m^{2}}{\Lambda^{2}},\frac{m^{2}}{\Lambda
^{2}}\ln\frac{m^{2}}{\Lambda^{2}})$ and $O(\frac{m^{2}}{M^{2}},\frac{m^{2}%
}{M^{2}}\ln\frac{M^{2}}{m^{2}})$ in the r.h.s. (\ref{ReR}), where
$\Lambda\lesssim M_{\rho}$ is the characteristic scale of the form factor
$G(t)$. The aim of this work is to improve the result (\ref{R0}) for the
amplitude $\mathcal{A}\left(  q^{2}=0\right)  $ of the $P\rightarrow
l^{+}l^{-}$ decay by taking into account all order mass corrections $\sim
\frac{m^{2}}{\Lambda^{2}},\frac{m^{2}}{\Lambda^{2}}\ln\frac{m^{2}}{\Lambda
^{2}}$.

\section{Mellin-Barnes integral representation}

We evaluate the amplitude $\mathcal{A}\left(  q^{2}\right)  $ following the
way used in \cite{Efimov:1981vh}. Let us transform the integral in (\ref{Rq})
to the Euclidean metric $k_{0}\rightarrow ik_{4}$. The corresponding integral
is convergent due to decreasing of $F_{P\gamma^{\ast}\gamma^{\ast}}%
(k^{2},(q-k)^{2})$ in the Euclidean region. Then use the double Mellin
transformation for the meson form factor%
\begin{equation}
F_{P\gamma^{\ast}\gamma^{\ast}}(k^{2},(q-k)^{2})=\frac{1}{\left(  2\pi
i\right)  ^{2}}\int_{\sigma+iR^{2}}dz\Phi\left(  z_{1},z_{2}\right)  \left(
\frac{\Lambda^{2}}{k^{2}}\right)  ^{z_{1}}\left(  \frac{\Lambda^{2}}{\left(
k-q\right)  ^{2}}\right)  ^{z_{2}}, \label{FM}%
\end{equation}
where $\Lambda$ is the characteristic scale for the form factor, $dz=$
$dz_{1}dz_{2}$, the vector $\sigma=\left(  \sigma_{1},\sigma_{2}\right)  \in\mathbb{R}
^{2},$ and $\Phi\left(  z_{1},z_{2}\right)  $ is the inverse Mellin transform
of the form factor
\begin{equation}
\Phi\left(  z_{1},z_{2}\right)  =\int_{0}^{\infty}dt_{1}\int_{0}^{\infty
}dt_{2}t_{1}^{z_{1}-1}t_{2}^{z_{2}-1}F_{P\gamma^{\ast}\gamma^{\ast}}\left(
t_{1},t_{2}\right)  \label{PhiM}%
\end{equation}
which has singularities at $\operatorname{Re}(z_{i})=0,-1,..$. . Introducing
Feynman parameters in the standard way, the denominator part of the integrand
in (\ref{Rq}) can be written as
\begin{align}
&  \frac{1}{\left(  k^{2}\right)  ^{z_{1}+1}\left[  \left(  k-q\right)
^{2}\right]  ^{z_{2}+1}\left[  (p_{-}-k)^{2}+m^{2}\right]  }\\
&  =\frac{\Gamma\left(  3+z_{1}+z_{2}\right)  }{\Gamma\left(  z_{1}+1\right)
\Gamma\left(  z_{2}+1\right)  }\int%
{\displaystyle\prod\limits_{i=1}^{3}}
d\alpha_{i}\delta\left(  1-\sum_{i=1}^{3}\alpha_{i}\right)  \frac{\alpha
_{1}^{z_{1}}\alpha_{2}^{z_{2}}}{\left[  k^{2}+D\right]  ^{3+z_{1}+z_{2}}%
},\nonumber
\end{align}
where $D=\left(  \alpha_{3}^{2}m^{2}-\alpha_{1}\alpha_{2}p^{2}\right)  $. Then
the $k-$loop integral reduces to%
\begin{equation}
\frac{2}{p^{2}}\int\frac{d^{4}k}{\pi^{2}}\frac{(pk)^{2}-p^{2}k^{2}}{\left[
k^{2}+D\right]  ^{3+z_{1}+z_{2}}}=\frac{\Gamma\left(  z_{1}+z_{2}\right)
}{\Gamma\left(  3+z_{1}+z_{2}\right)  D^{z_{1}+z_{2}}}\left[  -3+2\frac
{\alpha_{3}^{2}}{D}\left(  m^{2}-\frac{1}{4}p^{2}\right)  \left(  z_{1}%
+z_{2}\right)  \right]  .\nonumber
\end{equation}
Combining all factors we get%
\begin{align}
&  \mathcal{A}\left(  q^{2}\right)  =\frac{1}{\left(  2\pi i\right)  ^{2}}%
\int_{\sigma+iR^{2}}dz\frac{\Phi\left(  z_{1},z_{2}\right)  \left(
\Lambda^{2}\right)  ^{z_{1}+z_{2}}\Gamma\left(  z_{1}+z_{2}\right)  }%
{\Gamma\left(  z_{1}+1\right)  \Gamma\left(  z_{2}+1\right)  }\int%
{\displaystyle\prod\limits_{i=1}^{3}}
d\alpha_{i}\delta\left(  1-\sum_{i=1}^{3}\alpha_{i}\right) \label{Ap1}\\
&  \cdot\frac{\alpha_{1}^{z_{1}}\alpha_{2}^{z_{2}}}{\left(  \alpha_{3}%
^{2}m^{2}-\alpha_{1}\alpha_{2}q^{2}\right)  ^{z_{1}+z_{2}}}\left[
-3+2\frac{\alpha_{3}^{2}\left(  m^{2}-\frac{1}{4}q^{2}\right)  }{\alpha
_{3}^{2}m^{2}-\alpha_{1}\alpha_{2}q^{2}}\left(  z_{1}+z_{2}\right)  \right]
.\nonumber
\end{align}

\begin{figure}[th]
\includegraphics[width=11cm]{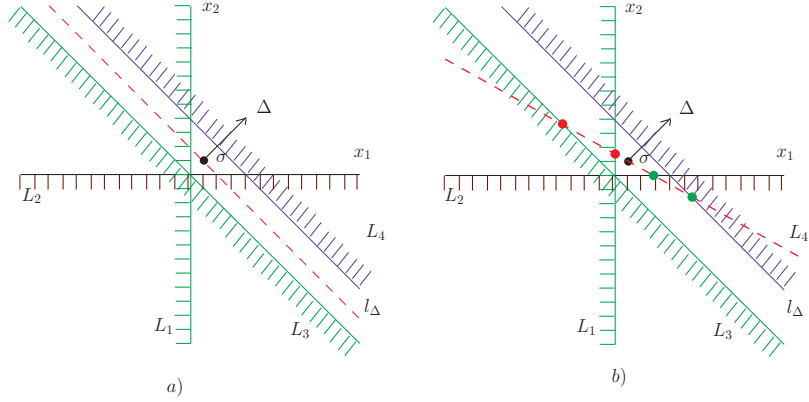}\caption{a) Singularities of the integrand in (\ref{A0}).
The semi-planes where the arguments of $\Gamma$ functions produce singularities are depicted
by lines {$\L_i$} with shadowed bands. The point $\sigma$ characterizing the integration counter
is from the triangle $\left\{x_{1}>0,x_{2}>0,x_{2}+x_{1}<\frac{1}{2}\right\}$
b) Rotation of $l_\Delta$ allows one to read off the degeneration.}
\label{plane}%
\end{figure}

In the general case, to step further we need to perform the third Mellin
transformation for denominators containing $\alpha_{i}$ \cite{Efimov:1981vh}.
Then, considering the process $P\rightarrow l^{+}l^{-}$ with mass hierarchy
$m<<M\leq\Lambda\sim m_{\rho}$ we expand the integral obtained over the ratios
of the lepton and meson masses to the characteristic scale of the meson form
factor $\Lambda$ by closing the Mellin contours in an appropriate manner.
However, in the present study we are interested in the amplitude at $q^{2}=0$.
In this limit$,$ the Feynman parameter integrals in (\ref{Ap1}) can be carried
out in terms of $\Gamma$-functions, and we obtain the following Mellin-Barnes
representation for $\mathcal{A}\left(  q^{2}=0\right)  $%
\begin{equation}
\mathcal{A}\left(  0\right)  =\frac{1}{\left(  2\pi i\right)  ^{2}}%
\int_{\sigma+i\mathbb{R}
^{2}}dz\left(  \xi^{2}\right)  ^{-z_{1}-z_{2}}\frac{\Gamma\left(
z_{1}\right)  \Gamma\left(  z_{2}\right)  \Gamma\left(  z_{12}\right)
\Gamma\left(  1-2z_{12}\right)  }{\Gamma\left(  3-z_{12}\right)  }\left[
\frac{\left(  -3+2z_{12}\right)  \Phi\left(  z_{1},z_{2}\right)  }%
{\Gamma\left(  z_{1}\right)  \Gamma\left(  z_{2}\right)  }\right]  ,
\label{A0}%
\end{equation}
with $\sigma$ in the triangle $\left\{  x_{1}>0,x_{2}>0,x_{2}+x_{1}<\frac
{1}{2}\right\}  $ chosen so that the integration path $\sigma+i\mathbb{R}
^{2}$ does not intersect the $\Gamma$-function singularities given by the
polar complex lines (see illustration in Fig. \ref{plane}a)
\begin{align}
L_{1}  &  :\left\{  z_{1}=-\nu\right\}  ,\quad L_{2}:\left\{  z_{2}%
=-\nu\right\}  ,\quad L_{3}:\left\{  z_{1}+z_{2}=-\nu\right\}  ,\label{L}\\
L_{4}  &  :\left\{  1-2\left(  z_{1}+z_{2}\right)  =-\nu\right\}  ,\qquad
\nu=0,1,2,...\nonumber
\end{align}
\qquad In (\ref{A0}) we introduce the notation $\xi^{2}=m^{2}/\Lambda^{2}$,
$z_{12}=z_{1}+z_{2}$ and combine the regular expression in the squared brackets.

In further analysis of the integral (\ref{A0}) we use the technique suggested
in \cite{Passare:1996db}.\footnote{When our study was completed we became
aware of the results of the work \cite{DeRafael:2008qj} where similar
technique of the two-dimensional counter integrals is used.}
Following this line let us associate the vectors in
2-dimensional space with the coefficients of the $\Gamma$-function arguments
in the numerator and denominator of the integrand in (\ref{A0}) $a_{1}=\left(
1,0\right), a_{2}=\left(  0,1\right), a_{3}=\left(  1,1\right),
a_{4}=\left(  -2,-2\right), c_{1}=\left(  -1,-1\right)$.
Next, define the vector
\begin{equation}
\Delta=\sum a_{i}-\sum c_{j}=\left(  1,1\right)
\end{equation}
and draw through $\sigma$ the straight line $l_{\Delta}=\left\{  x\in\mathbb{R}
^{2}:\left(  \Delta,x\right)  =\left(  \Delta,\sigma\right)  \right\}  $ with
the normal vector $\Delta$. The scalar product is introduced as $\left(
x,y\right)  =x_{1}y_{1}+x_{2}y_{2}$. The point $\sigma$ divides $l_{\Delta}$
into two rays $l^{+}$ and $l^{-}$ so that the pair of directions $l^{+}$ and
$\Delta$ yields the same orientation of $\mathbb{R}
^{2}$ as the pair of coordinate axes $x_{1}$ and $x_{2}$. The half-plane
$\pi_{\Delta}=\left\{  x\in\mathbb{R}
^{2}:\left(  \Delta,x\right)  <\left(  \Delta,\sigma\right)  \right\}  $ with
boundary $l_{\Delta}$ and the integration half-space $\Pi_{\Delta}=\pi
_{\Delta}+i\mathbb{R}
^{2}=\left\{  z\in C^{2}:\operatorname{Re}\left(  \Delta,z\right)
<\operatorname{Re}\left(  \Delta,\sigma\right)  \right\}  $ characterize the
domain in the space of integration variables $z$ in which the integrand is a
decreasing function.

Now we need to define the divisors given by the condition%
\begin{equation}
D_{1}=\cup\left\{  L_{j}:L_{j}\cap l^{-}=0\right\}  ,D_{2}=\cup\left\{
L_{j}:L_{j}\cap l^{+}=0\right\}  .
\end{equation}
The theorem \cite{Passare:1996db} states that in the nondegenerated case
($\Delta\neq0$ and all $a_{i}\nparallel\Delta$) the integral like (\ref{A0})
is given by the sum%
\begin{equation}
\mathcal{A}\left(  0\right)  =\sum_{z_{r}\in\Pi_{\Delta}}\underset{z_{r}}%
{res}\left[  \mathrm{Integ\operatorname{rand}}\mathcal{A}\left(  0\right)
\right]  ,
\end{equation}
where $\underset{z_{r}}{res}\left[  \mathrm{Integ\operatorname{rand}%
}\mathcal{A}\left(  0\right)  \right]  $ is the residue with respect to the
system of divisors $\left\{  D_{1},D_{2}\right\}  $.

The integral (\ref{A0}) corresponds to the degenerate case since $a_{3}$ and
$a_{4}$ are parallel to $\Delta$. In this case, one has $L_{2}\in D_{1}$ and
$L_{1}\in D_{2}$. However, $L_{3}$ and $L_{4}$ are parallel to $l_{\Delta}$
and cannot to be ascribed to any of divisors. To read off the degeneration we
slightly rotate $l_{\Delta}$ with respect to the point $\sigma$ in clock-wise
and anti-clock-wise directions (Fig. \ref{plane}b). Now we have crossings of $L_{3,4}$ with the
rotated line and are able to decide to which divisor they should be related.
One has two cases%
\begin{align}
1)D_{1}  &  =\left\{  L_{2},L_{4}\right\}  ,D_{2}=\left\{  L_{1}%
,L_{3}\right\}  ,\label{case}\\
2)D_{1}  &  =\left\{  L_{2},L_{3}\right\}  ,D_{2}=\left\{  L_{1}%
,L_{4}\right\}  .\nonumber
\end{align}

We are interested in the intersection of divisors, \textit{i.e.},
intersections of all $L_{i}^{(1)}\in D_{1}$ and all $L_{i}^{(2)}\in D_{2}$ so
that these intersections $L_{i}^{(1)}\cap L_{j}^{(2)}$ belong to the
half-space $\Pi_{\Delta}$.

Another important property is the intersection rank, the number of lines that
meet at each point. If only two lines $L^{(1)}$ and $L^{(2)}$ meet at each
point $z_{r}\in D_{1}\cap D_{2}\cap\Pi_{\Delta}$ (rank 1), then one has only
simple poles. If the intersection rank is more than one, than one may either
apply the theory of multiple residues or introduce small $\varepsilon
-$parameters in the arguments of $\Gamma-$functions in such way that all poles
become simple ones (like in the dimensional regularization method).

We prefer here the second approach, namely, $L_{1},L_{2},L_{3}$ in (\ref{L})
meet at the same points $(-\alpha,-\beta)$ where $\alpha,\beta=0,1,...$ are
positive integers. In order to get rid of this kind of degeneracy, we add a
small parameter $\varepsilon$ to the argument $\Gamma\left(  z_{2}\right)
\rightarrow\Gamma\left(  z_{2}+\varepsilon\right)  $ in (\ref{A0}). Now we
ready to analyze the poles and their residues. Consider first the case 1) in
(\ref{case}). We have two sets of intersections in $D_{1}\cap D_{2}\cap
\pi_{\Delta}$%
\begin{equation}
L_{2}\cap L_{1},L_{2}\cap L_{3}%
\label{cap}\end{equation}
which may be parametrized as%
\begin{equation}
L_{2}\cap L_{1}:\left\{
\begin{array}
[c]{c}%
z_{2}+\varepsilon=-\alpha,\\
z_{1}=-\beta,
\end{array}
\right.  ,\qquad L_{2}\cap L_{3}:\left\{
\begin{array}
[c]{c}%
z_{2}+\varepsilon=-\alpha,\\
z_{1}=\alpha-\beta+\varepsilon,
\end{array}
\right.  .
\end{equation}
Calculating residues we get two contributions to the integral%
\begin{equation}
\mathcal{A}\left(  0\right)  =\mathcal{A}_{a}\left(  0\right)  +\mathcal{A}%
_{b}\left(  0\right)
\end{equation}%
\begin{align}
\mathcal{A}_{a}\left(  0\right)   &  =\sum_{\alpha,\beta=0}^{\infty}%
\frac{\left(  -1\right)  ^{\alpha+\beta}}{\alpha!\beta!}\left(  \xi
^{2}\right)  ^{\alpha+\beta+\varepsilon}\frac{\Gamma\left(  -\alpha
-\beta-\varepsilon\right)  \Gamma\left(  1+2\left(  \alpha+\beta
+\varepsilon\right)  \right)  }{\Gamma\left(  3+\alpha+\beta+\varepsilon
\right)  }\label{Aa}\\
&  \cdot\left(  -3-2\left(  \alpha+\beta+\varepsilon\right)  \right)  \left[
\frac{\Phi\left(  -\alpha,-\beta\right)  }{\Gamma\left(  -\alpha\right)
\Gamma\left(  -\beta\right)  }\right]  ,\nonumber\\
\mathcal{A}_{b}\left(  0\right)   &  =\sum_{\alpha,\beta=0}^{\infty}%
\frac{\left(  -1\right)  ^{\alpha+\beta}}{\alpha!\beta!}\left(  \xi
^{2}\right)  ^{\beta}\frac{\Gamma\left(  1+2\beta\right)  }{\Gamma\left(
3+\beta\right)  }\left(  -3-2\beta\right)  \left[  \frac{\Phi\left(
\alpha-\beta+\varepsilon,-\alpha\right)  }{\Gamma\left(  -\alpha\right)
}\right]  . \label{Ab}%
\end{align}
The second case in (\ref{case}) is reduced to the first one because one has single parameter
($\xi$) integral. If one would be interested in the expansion in inverse powers of $\xi$
one needs to consider intersections $L_{4}\cap L_{1},L_{4}\cap L_{3}$ instead of (\ref{cap}).

By using the representation (\ref{PhiM}) one may show that the corresponding
residues are
\begin{equation}
\frac{\Phi\left(  -\alpha,-\beta\right)  }{\Gamma\left(  -\alpha\right)
\Gamma\left(  -\beta\right)  }=\left(  -1\right)  ^{\alpha+\beta}%
F_{P\gamma^{\ast}\gamma^{\ast}}^{\left(  \alpha,\beta\right)  }(0,0),\quad
\frac{\Phi\left(  z,-\alpha\right)  }{\Gamma\left(  -\alpha\right)  }=\left(
-1\right)  ^{\alpha}\int_{0}^{\infty}dtt^{z-1}F_{P\gamma^{\ast}\gamma^{\ast}%
}^{\left(  0,\alpha\right)  }\left(  t,0\right)  ,
\end{equation}
where $F_{P\gamma^{\ast}\gamma^{\ast}}^{\left(  \alpha,\beta\right)  }(0,0)$
denotes the derivatives of an order of $\alpha$ and $\beta$ in the
corresponding arguments of the form factor. After these substitutions one sum
in (\ref{Aa}) and (\ref{Ab}) may be performed with the result
\begin{align*}
&  \mathcal{A}_{a}\left(  0\right)  =-\sum_{n=0}^{\infty}\frac{G^{(n)}(0)}%
{n!}\left(  \xi^{2}\right)  ^{n+\varepsilon}\frac{\Gamma\left(  -n-\varepsilon
\right)  \Gamma\left(  1+2\left(  n+\varepsilon\right)  \right)  }%
{\Gamma\left(  3+n+\varepsilon\right)  }\left(  3+2\left(  n+\varepsilon
\right)  \right)  ,\\
&  \mathcal{A}_{b}\left(  0\right)  =\sum_{n=0}^{\infty}\frac{\left(
-1\right)  ^{n}}{n!}\left(  \xi^{2}\right)  ^{n}\frac{\Gamma\left(
1+2n\right)  \Gamma\left(  -\varepsilon\right)  }{\Gamma\left(  3+n\right)
\Gamma\left(  1-\varepsilon+n\right)  }\left(  3+2n\right)  \int_{0}^{\infty
}dtt^{\varepsilon}G^{\left(  n+1\right)  }\left(  t\right)  ,
\end{align*}
where we again use $G(t)\equiv F_{P\gamma^{\ast}\gamma^{\ast}}(t,t)$. Now we
expand in $\varepsilon$ and take the limit $\varepsilon\rightarrow0$ with the
total result%

\begin{align}
\mathcal{A}\left(  0\right)   &  =\sum_{n=0}^{\infty}\frac{\left(  -\xi
^{2}\right)  ^{n}}{n!}\frac{\Gamma\left(  1+2n\right)  }{\Gamma\left(
1+n\right)  \Gamma\left(  3+n\right)  }\left\{  G^{(n)}(0)\left[  2+\left(
3+2n\right)  \left(  \ln4\xi^{2}-\gamma\right.  \right.  \right. \label{A0S}\\
&  \left.  \left.  \left.  -\psi\left(  n+1\right)  +\psi\left(  n+\frac{1}%
{2}\right)  -\psi\left(  n+3\right)  \right)  \right]  +\left(  3+2n\right)
\int_{0}^{\infty}dtG^{\left(  n+1\right)  }\left(  t\right)  \ln t\right\}
.\nonumber
\end{align}
Note that the $\varepsilon^{-1}$ poles contained in the intermediate steps of
calculations are canceled in the final expression. To the lowest orders in
$\xi^{2}$ expansion one gets
\begin{align}
\mathcal{A}^{\left(  0\right)  }\left(  0\right)   &  =\frac{1}{2}\left[
3\ln\xi^{2}-\frac{5}{2}+3\int_{0}^{\infty}dtG^{\left(  1\right)  }\left(
t\right)  \ln t\right]  ,\label{A00}\\
\mathcal{A}^{\left(  1\right)  }\left(  0\right)   &  =-\xi^{2}\frac{1}%
{3}\left[  G^{(1)}(0)\left(  5\ln\xi^{2}+\frac{13}{6}\right)  +5\int
_{0}^{\infty}dtG^{\left(  2\right)  }\left(  t\right)  \ln t\right]  .
\label{A01}%
\end{align}
The leading order expression (\ref{A00}) is in accordance with the result
(\ref{R0}) obtained in \cite{Dorokhov:2007bd}. In the general case it is
convenient to convert the sum in (\ref{A0S}) into the integral form%
\begin{align}
&  \mathcal{A}\left(  0\right)  =\frac{4}{3\pi}\int_{0}^{1}dy\sqrt{\frac
{1-y}{y}}\left\{  \left[  \left(  \ln4\xi^{2}-\gamma\right)  \left(
2+y\right)  +2\left(  1-y\right)  \right]  G\left(  -4y\xi^{2}\right)
+\right. \nonumber\\
&  \left.  +\left(  2+y\right)  \int_{0}^{\infty}dt\left[  \ln tG^{\left(
1\right)  }\left(  t-4y\xi^{2}\right)  +\frac{e^{-\frac{1}{2}t}-e^{-3t}%
-e^{-t}}{e^{-t}-1}G\left(  -4ye^{-t}\xi^{2}\right)  -\frac{e^{-t}}{t}G\left(
-4y\xi^{2}\right)  \right]  \right\}  . \label{A0I}%
\end{align}

Finally, let us consider the form factor we are interested in from a physical
point of view%
\begin{equation}
G\left(  t\right)  =\frac{1}{1+t}.\nonumber
\end{equation}
For this form factor from (\ref{A00}) - (\ref{A0I}) one gets the coefficient
of logarithmic term as%
\begin{equation}
\mathcal{A}\left(  0\right)  =\frac{1}{12\xi^{4}}\left[  1+6\xi^{2}%
-\sqrt{1-4\xi^{2}}\left(  1+8\xi^{2}\right)  \right]  \ln\xi^{2}+O\left(
\xi^{0}\right)  ,
\end{equation}
or the first terms of expansion
\begin{equation}
\mathcal{A}\left(  0\right)  =\frac{3}{2}\left(  1+\frac{10}{9}\xi
^{2}+O\left(  \xi^{4}\right)  \right)  \ln\xi^{2}-\frac{5}{4}\left(
1+\frac{86}{45}\xi^{2}+O\left(  \xi^{4}\right)  \right)  .
\end{equation}
Thus, one can see that in the realistic case for muon, $\xi^{2}=m_{\mu}%
^{2}/\Lambda^{2}\sim m_{\mu}^{2}/m_{\rho}^{2}\approx0.02$ the corrections to
the leading order coefficients are of an order of $1\%$ and for an electron
pair they are negligible.

\section{Conclusions}

The aim of this paper is to clarify the situation with rare decays of
pseudoscalar mesons to a lepton pair. The situation became more pressing after
recent KTeV E799-II experiment at Fermilab in which the pion decay into an
electron-positron pair was measured using the $K_{L}\rightarrow3\pi$ process
as a source of tagged neutral pions \cite{Abouzaid:2007md}. The branching
ratio was determined to be equal to
\begin{equation}
B_{\mathrm{no-rad}}^{\mathrm{KTeV}}\left(  \pi^{0}\rightarrow e^{+}%
e^{-}\right)  =\left(  7.49\pm0.29\pm0.25\right)  \cdot10^{-8}. \label{KTeV}%
\end{equation}
The standard model prediction based on the use of CLEO data on the transition
form factor $\pi\rightarrow\gamma\gamma^{\ast}$ \cite{Gronberg:1997fj} gives
\cite{Dorokhov:2007bd}
\begin{equation}
B^{\mathrm{Theor}}\left(  \pi^{0}\rightarrow e^{+}e^{-}\right)  =\left(
6.2\pm0.1\right)  \cdot10^{-8}, \label{Bth}%
\end{equation}
which is $3.3\sigma$ below the KTeV result (\ref{KTeV}). Therefore, it is
extremely important to trace possible sources of the discrepancy between the
experiment and theory. There are a number of possibilities: (1) problems with
(statistic) experiment procession, (2) inclusion of QED radiation corrections
by KTeV is wrong, (3) unaccounted mass corrections are important, and (4)
effects of new physics. At the moment the last possibilities was
reinvestigated. In \cite{Dorokhov:2008qn}, the contribution of QED radiative
corrections to the $\pi^{0}\rightarrow e^{+}e^{-}$ decay, which must be taken
into account when comparing the theoretical prediction (\ref{Bth}) with the
experimental result, (\ref{KTeV}) was revised. Comparing with earlier
calculations \cite{Bergstrom:1983ay}, the main progress is in the detailed
consideration of the $\gamma^{\ast}\gamma^{\ast}\rightarrow e^{+}e^{-}$
subprocess and revealing of dynamics of large and small distances.
Occasionally, this number agrees well with the earlier prediction based on
calculations \cite{Bergstrom:1983ay} and, thus, the KTeV analysis of radiative
corrections is confirmed. In the present paper, we show that the mass
corrections are under control and do not resolve the problem. So our main
conclusion is that the inclusion of radiative and mass corrections is unable
to reduce the discrepancy between the theoretical prediction for the decay
rate (\ref{Bth}) and experimental result (\ref{KTeV}). The effects of new
physics were considered in \cite{Kahn:2007ru} where the excess of experimental
data over theory is explained by the contribution of low mass ($\sim10$ MeV)
vector bosons appearing in some models of dark matter. Further independent
experiment at KLOE, NA48, WASAatCOSY, BES III and other facilities will be crucial for
resolution of the problem.

\section{Acknowledgments}

We are grateful to E.A. Kuraev, N. I. Kochelev, A.V. Kotikov, and S.V.
Mikhailov for helpful discussions on the subject of this work. A.E.D.
acknowledges partial support from the JINR-INFN program and the Scientific
School grant 4476.2006.2.

\end{document}